# Coherently averaged dual-comb spectroscopy with a low-noise and high-power free-running gigahertz dual-comb laser

C. R. Phillips,[1,*] B. Willenberg,[1] A. Nussbaum-Lapping,[1] F. Callegari,[2,3] S. L. Camenzind,[1] J. Pupeikis,[1] and U. Keller[1]

[1] *Department of Physics, Institute for Quantum Electronics, ETH Zurich, Switzerland*
[2] *Nanoscopy and NIC@IIT, Istituto Italiano di Tecnologia (IIT), Genova, Italy*
[3] *Department of Physics, University of Genova, Genova, Italy*
[*] *cphillips@phys.ethz.ch*

**Abstract:** We present a new type of dual optical frequency comb source capable of scaling applications to high measurement speeds while combining high average power, ultra-low noise operation, and a compact setup. Our approach is based on a diode-pumped solid-state laser cavity which includes an intracavity biprism operated at Brewster angle to generate two spatially-separated modes with highly correlated properties. The 15-cm-long cavity uses an Yb:CALGO crystal and a SESAM as an end mirror to generate more than 3 W average power per comb, below 80 fs pulse duration, a repetition rate of 1.03 GHz, and a continuously tunable repetition rate difference up to 27 kHz. We carefully investigate the coherence properties of the dual-comb by a series of heterodyne measurements, revealing several important features: (1) ultra-low jitter on the uncorrelated part of the timing noise; (2) the radio frequency comb lines of the interferograms are fully resolved in free-running operation; (3) we validate that through a simple measurement of the interferograms we can determine the fluctuations of the phase of all the radio frequency comb lines; (4) this phase information is used in a post-processing routine to perform coherently averaged dual-comb spectroscopy of acetylene ($C_2H_2$) over long timescales. Our results represent a powerful and general approach to dual-comb applications by combining low noise and high power operation directly from a highly compact laser oscillator.

## 1. Introduction

The introduction of optical frequency comb (OFC) technology has had a major impact on laser-based measurements [1–3]. Dual-comb sources, consisting of a pair of OFCs, are a particularly useful configuration [4, 5]. Since the two combs have a small repetition rate difference, the delay between the pulses is swept over time. This sweeping can replace mechanical delay line based measurements, offering longer range and/or higher scan speeds without moving parts. The applications of dual-comb sources are diverse, including high-resolution spectroscopy of trace gases, laser ranging, and pump-probe measurements.

A large amount of work on dual-comb lasers and their applications has involved sources with pulse repetition rates $f_{\text{rep}}$ of order 100 MHz, with corresponding delay ranges $1/f_{\text{rep}}$ of around 10 ns. Such sources enable high resolution gas spectroscopy [6, 7], or pump-probe measurements on complicated thin films comprised of many layers [8]. However, scaling to gigahertz lasers is also appealing for several reasons. For example, a repetition rate of 1 GHz offers fast pump-probe sampling [9, 10], provides a sufficient density of comb lines to directly resolve absorption features in condensed matter targets or gases at ambient pressures [11–14], and still supports long-distance dual-comb laser ranging through use of the Vernier effect [15, 16]. Gigahertz lasers can typically measure signals much faster than their megahertz counterparts due to faster update rates and higher sensitivity, enabling real-time measurements and observation of non-repeating signals.

To address such applications it is important for the optical system to be sufficiently simple and robust. This goal has motivated extensive work on single-cavity dual-comb lasers, in which both combs share the same cavity optics. The motivation for this approach is two-fold: the influence of noise is reduced due to correlated fluctuations on the lasers' timing and phase fluctuations, and the laser system is simplified since only one free-running cavity is needed instead of two cavities with various fast active feedback control loops. Single-cavity dual-comb generation has been shown for the first time in semiconductor disk lasers [17], bulk crystal lasers [18], and fiber lasers [19]. In general, solid state lasers are favorable in terms of noise properties due to their low-loss, low-dispersion, low-nonlinearity cavities [20, 21]. In recent years, dual-comb modelocking has been extended to femtosecond solid-state lasers [22–26]. In our group, we have recently introduced a complementary approach based on spatial multiplexing using a Fresnel biprism, demonstrated via an 80-MHz laser [27]. This spatial multiplexing approach splits the cavity mode into two similar paths with small but finite beam separation on the active intracavity components.

This technique is particularly beneficial when using birefringent solid-state gain materials having non-isotropic gain properties. Since both combs can have the same polarization state, it is relatively straightforward to obtain near-identical comb properties. Yb-doped materials are favorable for efficient high power femtosecond pulse generation, and among these Yb:CALGO is one of the most promising [28]. This birefingent material supports multiple watt average powers, sub-100-fs pulse durations, and high repetition rates [29, 30]. Recent work showed that gigahertz operation at high power can be obtained by Kerr-lens modelocking (KLM) as well, with sub-50-fs pulses [31].

In this work, we extend the spatial multiplexing approach to the gigahertz regime via a dual-comb modelocked Yb:CALGO laser. We perform a detailed noise analysis showing low noise operation with excellent coherence between the two combs in free running operation and thereby obtain comb-line-resolved measurements directly from the dual-comb interferograms (i.e. the interferometric signal obtained by combining the two combs at a beam splitter). Our approach uses a multimode diode pumped solid state laser configuration and a semiconductor saturable absorber mirror (SESAM) for modelocking [32]. The laser (presented in Section 2) generates > 3 W average power per comb with 78-fs pulses at 1.0327-GHz repetition rate around 1057-nm center wavelength, with up to 27 kHz repetition rate difference.

With free-running combs there are fluctuations in the comb properties that, if not properly accounted for, can compromise the measurements. This issue has motivated several studies in the community based on computational averaging and comb tracking [33–37]. Here we perform a detailed analysis on these aspects for our laser system in Sections 3 and 4. For practical applications it is beneficial to work purely with interferograms without requiring any auxillary lasers, Bragg gratings, or other components. To validate this is feasible with our laser while preserving comb line resolution (as needed for applications such as high-resolution spectroscopy), we first explain an approach to extrapolate the phase from the interferograms, and validate it by comparison to an independent measurement that uses an auxillary cw laser to fully track the phase of one radio frequency (RF) comb line. We thereby prove experimentally that the time-varying phase of any RF comb line can be predicted through the phase and timing information obtained from the interferograms. Therefore, all fluctuations of the RF comb lines of this free-running laser can be tracked and corrected for. By using this phase extrapolation for phase correction of the interferograms, we obtain coherent averaging over long timescales (Section 5). The high average power, high power per comb line, high speed, low noise, and compact nature of the laser make it a compelling source for a wide variety of sensing applications, including coherently averaged dual-comb spectroscopy which we demonstrate in a proof-of-principle experiment using an acetylene cell.

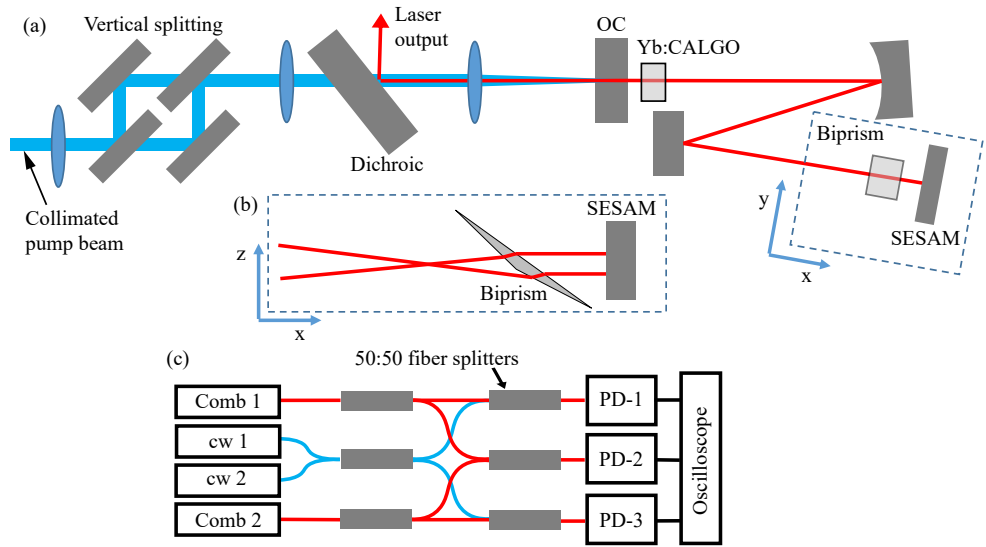

Fig. 1. Laser setup and diagnostics. (a) Top-down schematic of laser cavity and pump arrangement. The combs are multiplexed in the vertical, so from this view only one beam path is visible for the laser. A single pump diode is collimated and split into two parts to pump each comb separately as shown on the left of the figure. (b) Side-view schematic around the SESAM end of the cavity, showing how the combs are multiplexed at the Brewster-angled biprism. The combs are also separated vertically in the gain crystal. (c) Heterodyne measurement setup from [26]. Heterodyne measurements are generated between the two combs, and between the individual combs and the two cw lasers. Gigahertz-bandwidth photodiodes (PDs) are used for each signal.

## 2. Dual-comb laser

### 2.1. Laser cavity

We use a diode-pumped solid-state laser configuration to generate the gigahertz dual-comb source. Both combs are generated in a single cavity arrangement via spatial multiplexing. In our recent work, a Fresnel biprism was operated in reflection to split the cavity mode into two closed paths [27]. Here we use a complementary configuration, where the biprism is operated in transmission at Brewster angle. This configuration proved convenient given the restricted number and separation between optics inside a gigahertz cavity as compared to an 80-MHz cavity. The laser setup is depicted in Figure 1.

The cavity uses an end-pumped configuration where the two combs are parallel and spatially separated at the flat end mirror. The laser is pumped with a wavelength-stabilized diode operating at 980 nm and delivered through a 100-µm-diameter multimode fiber with an NA of 0.15. After collimation in free space, the pump light is split into two equal parts with a 50:50 beam splitter to provide one pump beam for each comb. These two pump beam components are routed into the gain crystal with imaging optics (see Fig. 1). The gain medium is an a-cut 2.5%-doped Yb:CALGO crystal (EOT) which is antireflection coated for the pump and laser wavelengths and has a length of 1.5 mm. The a-axis is oriented in the vertical direction. The two pump beam components are separated in the vertical direction in the gain crystal in order to overlap with the corresponding positions of the two combs.

The end mirror acts as an output coupler (2.6% transmission), has high transmission for the pump, and provides a negative group delay dispersion (GDD) of -600 fs$^2$ for the laser. The output laser beams are separated from the incoming pump beams via a 45° dichroic mirror and routed to beam diagnostics. As well as the dichroic output coupler, the cavity contains a 75-mm concave radius of curvature highly reflective mirror with low dispersion, a flat dispersion-compensating mirror (-100 fs$^2$ per reflection), a UV fused silica Brewster-oriented biprism (179° apex angle), and a semiconductor saturable absorber mirror (SESAM). The in-house grown SESAM has saturation fluence 21 µJ/cm$^2$, and a modulation depth of 1.2%.

The intracavity biprism is oriented such that the Brewster condition is obtained for vertically polarized light and the beams transmitted through the two facets are deflected vertically [see Figure 1(b)]. By placement of the biprism directly in front of the SESAM, the cavity supports a pair of modes that are vertically polarized and well separated in the vertical direction on the gain crystal and on the SESAM. The repetition rate difference is adjusted by translation of the biprism up and down, since this varies the relative path length seen by the two combs. The biprism is mounted on a one-axis translation stage for this purpose. The limit to the tuning range is set by clipping of one or the other beam on the apex of the biprism; until such clipping occurs, the biprism translation does not perturb laser operation or the output beam path. The cavity uses a standard breadboard optomechanical arrangement with pedestals, adjustable mirror mounts, and is enclosed in a plastic box. The pump diode, laser crystal, and SESAM are temperature stabilized via Peltier cooling. The heat is extracted by water for the pump diode and laser crystal, and by convection for the SESAM.

*2.2. Laser characterization*

To characterize the laser, we measure modelocking diagnostics and then analyze various noise properties. The modelocking diagnostics consist of the optical spectrum analyzer (OSA), microwave spectrum analyzer (MSA), second-harmonic generation (SHG) autocorrelation (AC), and a signal-source analyzer (SSA). The laser timing jitter is inferred via measuring the electronic phase noise of the 15$^{\text{th}}$ harmonic of the repetition rate with the SSA (Keysight, E5052B). This harmonic is obtained by measuring the pulse train with a fast photodiode followed by an electronic band-pass filter. To characterize the dual-comb noise properties, we use the fiber-based heterodyne measurement setup described in [26]. This setup is conceptually illustrated in Figure 1(c) and described in Section 3.

Both combs exhibit simultaneous self-starting modelocking behavior with very similar properties of each beam. We obtain up to 3.07 W (comb 1) and 3.04 W (comb 2) with a total pump power of 18.5 W. The temporal and spectral characteristics are very similar, as shown in Figure 2. In particular, the combs have center wavelength of 1057 nm (1056 nm) and full-width at half maximum (FWHM) pulse duration of 78 fs as inferred from SHG autocorrelation. The nominal repetition rate is 1.0327 GHz. The repetition rate difference can be tuned from zero up to approximately 27 kHz without introducing any observable losses on the biprism.

## 3. Beat note setup theory

A crucial metric for dual-comb sources is their coherence, which we evaluate using the heterodyne measurement setup illustrated in Figure 1(c). In this section we present the relevant theoretical aspects of these measurements and then in Section 4 we present the experimental results.

When the pulses from the two combs arrive at around the same time at a beam splitter, the pulses interfere. Such temporal overlaps occur at a rate of $\Delta f_{\text{rep}}$. Hence, fluctuations in the relative phase and timing between the two combs is sampled at sampling intervals of $1/\Delta f_{\text{rep}}$. This infrequent sampling is not always enough to resolve the underlying fluctuations if the noise is too high. The heterodyne measurement setup overcomes this limitation by incorporating measurements with a pair of single-frequency lasers. Since these cw lasers are "always on" it becomes possible to track both fast and slow phase changes. This setup is described in detail in [26] and summarized in the following.

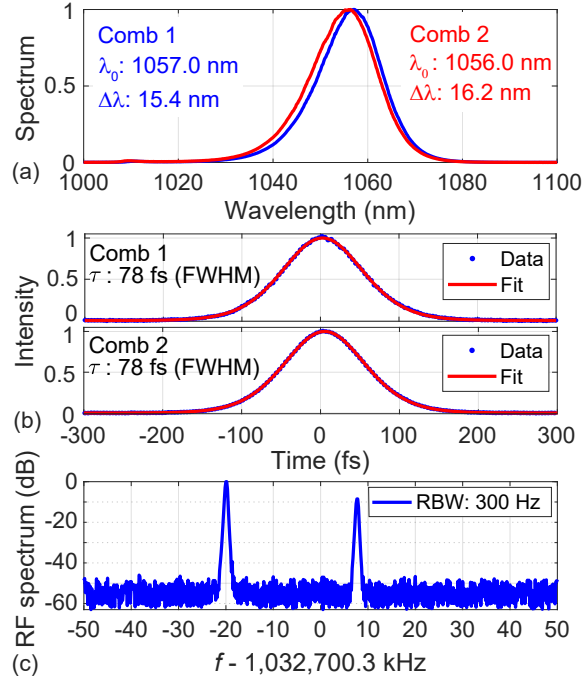

Fig. 2. Modelocking diagnostics of the dual-comb oscillator. (a) Optical spectrum of the two combs. (b) SHG autocorrelation of the two combs and corresponding sech$^2$ fits. (c) Microwave spectrum, zoomed around the first repetition rate harmonic; note that both combs are incident on the photodiode simultaneously in order to show the repetition rate difference. RBW: resolution bandwidth.

### 3.1. Overview of heterodyne measurement setup

In the measurement setup [Fig. 1(c)], two continuous wave (cw) lasers are combined interferometrically with the two combs to yield heterodyne beat note signals, and the two combs are combined to yield interferograms (IGMs). The setup thus yields four heterodyne beat notes that are captured simultaneously in two oscilloscope channels (via PD-1 and PD-3), corresponding to beating between each cw laser and a line of each comb. The RF bandwidth occupied by the signals is relatively narrow, so it is straightforward to ensure that the two beat notes that are on each channel are not overlapped in Fourier space. The individual beat note signals can therefore be extracted from the saved oscilloscope traces by digital band-pass filters. For computational efficiency of the subsequent processing steps, this filtering is implemented via a downsampling approach. For each signal, we perform a manual peak search in frequency space, shift the signal to DC, low-pass filter it (width of several MHz), and downsample onto a new time grid. The original frequency of each beat note signal (before shifting to DC) is kept track of so that no information is lost in these calculations.

The rate of phase change of the beat note signals is given, up to a sign flip, by

$$F_{\text{cw}-j}^{\text{comb}-i} = \frac{1}{2\pi}\frac{d}{dt}\left(\varphi_{\text{cw}-j}^{\text{comb}-i}\right)$$
$$= f_{\text{CEO}}^{\text{comb}-i} + N_{\text{cw}-j}^{\text{comb}-i} f_{\text{rep}}^{\text{comb}-i} - \nu_{\text{cw}-j}, \quad (1)$$

where superscript "comb $-$ i" denotes a quantity associated with comb $i \in \{1, 2\}$ and subscript "cw $-$ j" denotes a quantity associated with cw laser $j \in \{1, 2\}$. $\varphi$ is the phase of the heterodyne beatnote signal, $f_{\text{CEO}}$ and $f_{\text{rep}}$ are comb properties (carrier envelope offset frequency and repetition rate), $v_{\text{cw}-j}$ is the optical frequency of cw laser $j$, and $N_{\text{cw}-j}^{\text{comb}-i}$ is the comb line index of comb $i$ giving rise to the beat note signal with cw laser $j$. By taking suitable products of the heterodyne signals (with conjugation of one signal if needed to switch between a frequency sum or a frequency difference), we obtain signals whose oscillation frequency corresponds to differences between the terms from Eq. (1). In particular,

$$\begin{aligned}\Delta F_{\text{cw}-j} &= F_{\text{cw}-j}^{\text{comb}-2} - F_{\text{cw}-j}^{\text{comb}-1} \\ &= \Delta f_{\text{CEO}} + n_{\text{cw}-j}\Delta f_{\text{rep}} + m_{\text{cw}-j}f_{\text{rep}}^{\text{comb}-1},\end{aligned} \quad (2)$$

where $n_{\text{cw}-j} = N_{\text{cw}-j}^{\text{comb}-2}$, and $m_{\text{cw}-j} = N_{\text{cw}-j}^{\text{comb}-2} - N_{\text{cw}-j}^{\text{comb}-1}$. The carrier envelope offset difference is $\Delta f_{\text{CEO}} = f_{\text{CEO}}^{\text{comb}-2} - f_{\text{CEO}}^{\text{comb}-1}$, and the repetition rate difference is $\Delta f_{\text{rep}} = f_{\text{rep}}^{\text{comb}-2} - f_{\text{rep}}^{\text{comb}-1}$. Note, the frequencies $\Delta F_{\text{cw}-j}$ correspond to the frequency of a particular RF comb line of the interferograms, which is how the setup gives direct access to the RF comb line phase. The factor $m_{\text{cw}-j}$ is non-zero when the optical comb lines nearest to each other have different comb line indices, which is naturally the case when $\Delta f_{\text{rep}}$ is large. Provided the dual-comb interferograms are not aliased and the cw lasers lie within the optical bandwidth of the pulses, we can generally assume $m_{\text{cw}-1} = m_{\text{cw}-2}$.

By taking another layer of products between the beatnote signals (again with suitable conjugation of one signal if needed), we obtain a signal whose oscillation frequency is proportional to the repetition rate difference:

$$\begin{aligned}F_D &= \Delta F_{\text{cw}-2} - \Delta F_{\text{cw}-1} \\ &= (n_{\text{cw}-2} - n_{\text{cw}-1})\Delta f_{\text{rep}},\end{aligned} \quad (3)$$

where we have assumed $m_{\text{cw}-1} = m_{\text{cw}-2}$. Since the frequency $F_D$ is proportional to $\Delta f_{\text{rep}}$, the corresponding time-varying phase captures the phase noise associated with uncorrelated timing jitter between the two combs. For convenience, we denote this phase as $\varphi_D(t)$, with the following relation between phase and frequency: $d(\varphi_D)/dt = 2\pi F_D(t)$.

Since the absolute frequencies of the cw lasers are not known precisely, the index difference in Eq. (3) is not known before the measurement and must be inferred by other means. The interferograms can be used for this purpose as follows. A peak of the envelope of the interferogram corresponds to a time when the optical pulses are temporally overlapped, which in turn implies that all the RF comb lines are in phase with each other, and hence that any signal oscillating at frequency $\Delta f_{\text{rep}}$ has accumulated $2\pi$ phase (even if $\Delta f_{\text{rep}}$ is varying with time). Hence, we can write

$$n_{\text{cw}-2} - n_{\text{cw}-1} = \frac{\varphi_D(t_{\text{pk},j+k}) - \varphi_D(t_{\text{pk},j})}{2\pi k}, \quad (4)$$

where $t_{\text{pk},j}$ denotes the temporal center of interferogram $j$ and the integer $k \geq 1$. For the data taken for this paper, the direct calculation of the right hand side of this equation is very close to an integer ($\ll 10^{-6}$ difference), and so the value is rounded to obtain $n_{\text{cw}-2} - n_{\text{cw}-1}$. Knowing the index difference, we obtain an accurate and high-resolution measurement of the time-dependent repetition rate difference $\Delta f_{\text{rep}} = F_D(t)/(n_{\text{cw}-2} - n_{\text{cw}-1})$. Since the underlying frequency $F_D(t)$ is in the megahertz range, fluctuations at frequencies exceeding $\Delta f_{\text{rep}}$ can be tracked.

### 3.2. Interferogram structure

In this subsection, we consider the structure of the interferograms themselves. In general, we obtain interferograms by coherently combining the two combs and applying a digital low-pass filter with a cut-off at $f_{\text{rep}}/2$. To avoid aliasing effects, the corresponding radio frequency (RF) spectrum is centered near $f_{\text{rep}}/4$ by fine-tuning $\Delta f_{\text{rep}}$. A typical experimental signal is illustrated in Figure 3.

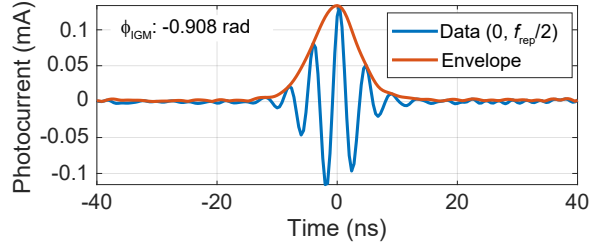

Fig. 3. Example dual-comb interferogram (IGM). The data from the oscilloscope has been converted to units of photocurrent, and then digitally low-passed (with a cut-off at $f_{\text{rep}}/2$) in order to show the dual-comb signal. The carrier envelope phase $\phi_{\text{IGM}}$ of the trace is shown in the top left of the figure. To obtain $\phi_{\text{IGM},j}$ for each IGM period $j$, first the temporal center $t_{\text{pk},j}$ is determined, and then the time-varying phase of the trace is evaluated at $t_{\text{pk},j}$.

With an ideal (noise-free) source, these interferograms (IGMs) are perfectly periodic up to an overall phase slip from one period to the next, in analogy to the carrier envelope offset frequency for individual optical frequency combs. The period is given by $\Delta f_{\text{rep}}$. The IGMs can also be viewed as a sum over all RF comb lines, and the time-dependent phase of one line is, following Eq. (2),

$$\phi_n(t) = 2\pi \left( \Delta f_{\text{CEO}} + n\Delta f_{\text{rep}} + m f_{\text{rep}}^{\text{comb}-1} \right) t \qquad (5)$$

for integers $n$ and $m$. Here we have assumed zero phase at $t = 0$ for simplicity. Given this equation, the phase slip from one period to the next is found by setting $t = 1/\Delta f_{\text{rep}}$:

$$\Delta\phi_n = 2\pi \left( \frac{\Delta f_{\text{CEO}} + m f_{\text{rep}}^{\text{comb}-1}}{\Delta f_{\text{rep}}} + n \right). \qquad (6)$$

Dual-comb measurements often require averaging multiple periods to improve the signal to noise ratio. Such coherent averaging requires that the quantity involved is perfectly periodic, so phase and timing fluctuations must be removed before averaging. If we consider a real-valued IGM denoted $V_{\text{IGM}}(t)$, then by suppressing the negative Fourier frequency components and applying a scale factor we can obtain an analytic signal $V_+(t)$ which satisfies $V_{\text{IGM}}(t) = \text{Re}[V_+(t)]$. To obtain a (nearly) periodic quantity, we subtract the phase in Eq. (5) from $V_+(t)$:

$$U_n(t) = V_+(t) \exp\left(-i\phi_n(t)\right). \qquad (7)$$

This new quantity has a comb structure with comb lines at integer multiples of $\Delta f_{\text{rep}}$ for any choice of $n$. Hence, all periods are in phase with each other and can be coherently summed. Coherent averaging is a computationally efficient way to perform dual-comb spectroscopy [38]. In Section 5 we describe the specific approach for coherent averaging used here to perform dual-comb spectroscopy on acetylene and compare it to the approach of [36].

The main challenge in successful averaging, especially for free-running combs, is determining $\phi_n(t)$. This quantity can fluctuate significantly, but it is only sampled infrequently (i.e. once per period of the IGM). The heterodyne measurement setup solves this problem by directly yielding $\phi_n(t)$ at all times between the interferogram peaks. However, for developing dual-comb measurement applications it is beneficial (in terms of system cost and complexity) to avoid using additional lasers or setups. In such situations, $\phi_n(t)$ must be determined via the IGMs themselves. This task can be readily accomplished with our new gigahertz dual-comb laser due to its low noise characteristics, which are discussed in the next section.

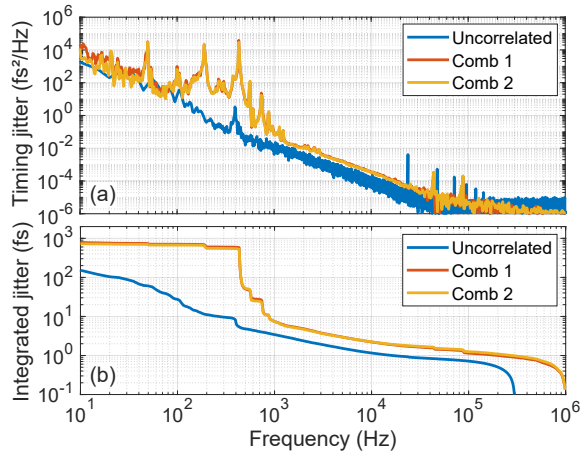

Fig. 4. Timing jitter measurements. (a) Comparison between the uncorrelated timing jitter [as inferred via the heterodyne measurement setup of Figure 1(c)] and the timing jitter of the individual combs [as inferred via the 15$^{\text{th}}$ harmonic of the individual repetition rates measured with the SSA]. Note, the high frequency peaks in the uncorrelated curve are at multiples of $\Delta f_{\text{rep}}$, which was 24 kHz for this measurement. Hence, "uncorrelated" refers to the timing jitter power spectral density (TJ-PSD) of $\Delta f_{\text{rep}}$ and "comb j" refers to the TJ-PSD of $f_{\text{rep}}^{\text{comb}-\text{j}}$. (b) Corresponding integrated timing jitter of the one-sided power spectral densities in (a).

## 4. Noise measurements

In this section, we analyze the noise properties of the dual-comb source. Except for SSA measurements, all other analysis uses a single set of traces from the heterodyne measurement setup that is representative of laser performance in the lab environment.

### 4.1. Timing jitter

The timing jitter measurements follow the calculations of Section 3.1. As with our other recent dual-comb results [26, 27], we compare the uncorrelated timing noise with the noise of the individual combs to see the extent by which the uncorrelated part has been suppressed by having the combs share the same pump and cavity. The results, shown in Figure 4, reveal a high amount of noise suppression over most of the range. The noise floor for the uncorrelated part is reached at slightly below 100 kHz, and this noise floor dominates the integrated jitter for frequencies above 30 kHz. The integrated noise is low, reaching 3 fs at 1 kHz integration limit. The noise is suppressed by at least 10 dB from 100 Hz up until the noise floor of the heterodyne measurement. The noise curves have an approximately $1/f^2$ slope in the kHz frequency range, which is consistent with relative intensity noise induced timing noise sources [21, 39]. There is a noticeable increase in the noise at sub-kHz frequencies, likely attributable to mechanical resonances in the optomechanics. Nonetheless, the dominant mechanical noise sources in the 100 Hz to 1 kHz range are strongly suppressed.

## 4.2. Interferogram phase tracking

Next we analyze the phase fluctuations of the IGMs. As described by Eq. (6), IGMs experience a phase slip from one period to the next. Given the presence of noise, the value of this phase slip varies over time. For each IGM period we obtain a phase $\phi_{\text{IGM},j}$, which corresponds to a sampling of the phase at the temporal center $t_{\text{pk},j}$. The individual IGM phase samples do not suffer from excessive noise since each IGM is well resolved in the time domain. However, the fluctuations from one IGM to the next are typically undersampled. Therefore, a phase unwrapping routine is required to recover an array of absolute IGM phases, which we denote $\boldsymbol{\phi}_{\text{IGM}}$. Note, symbol $\phi$ indicates quantities associated with the IGM phase or the RF comb lines; $\varphi$ indicates other phases derived from the heterodyne measurements [see Eqs. (1) and (4)]; and bold symbols indicate an array. The derivatives of $\boldsymbol{\phi}_{\text{IGM}}$, found by finite differences of adjacent sample points, are denoted $\Delta(\boldsymbol{\phi}_{\text{IGM}})$ (first-order derivative) and $\Delta(\Delta(\boldsymbol{\phi}_{\text{IGM}}))$ (second-order derivative). The unwrapped phase array $\boldsymbol{\phi}_{\text{IGM}}$, which is shown in Figure 5 for a typical measurement, was obtained as follows:

1. A small part of the trace around each IGM peak is analyzed to determine the carrier envelope phase of the electronic waveform for that period. This quantity corresponds to $\mathrm{mod}_{2\pi}(\phi_{\text{IGM},j})$ for period $j$.
2. The difference between the (wrapped) phase from one IGM to the next is taken, and a standard phase unwrap function is applied to the corresponding array. This calculation yields $\Delta(\boldsymbol{\phi}_{\text{IGM}})$ and the result is shown as the blue curve in Figure 5(a).
3. The unwrapped phase difference is cumulatively summed to arrive at the unwrapped IGM phase itself, i.e. $\boldsymbol{\phi}_{\text{IGM}}$. The result is as shown by the red curve in Figure 5(a).

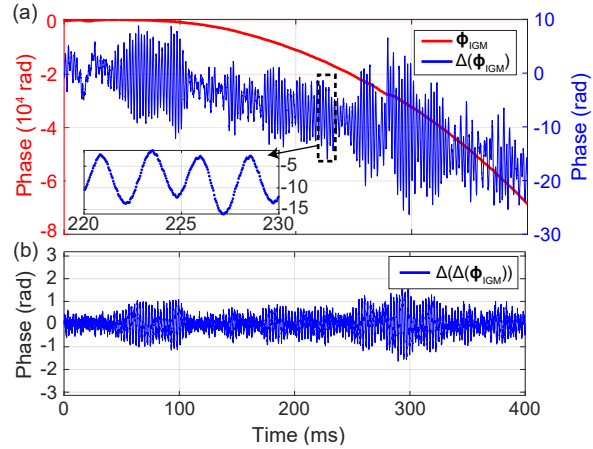

Fig. 5. Interferogram phase as a function of time. (a) The unwrapped phase (red) and its derivative (blue). The inset shows the behavior in the region indicated by the dashed black box, highlighting the influence of sub-kHz perturbations likely originating from mechanical noise. For this inset, the individual phase samples are shown as dots to illustrate that the oscillations are well sampled. (b) second derivative of the phase, illustrating that the fluctuations stay well within $\pm\pi$.

In this procedure, unwrapping of the phase difference rather than the phase itself is necessary due to the large fluctuations involved. For example, the phase difference curve $\Delta(\boldsymbol{\phi}_{\text{IGM}})$ varies by more than $2\pi$, which would be problematic when trying to unwrap the IGM phase directly. A more quantitative picture of this issue can be obtained from Figure 5(b), which shows the second-order difference $\Delta(\Delta(\phi_{IGM}))$. Since all elements of this array are well within $\pm\pi$, it is clear that there were no $2\pi$ phase ambiguities in the determination of the first-order difference, and hence the unwrapping procedure was successful. Here we used $\Delta(\phi_{IGM})$ for unwrapping. It is worth noting that in cases where this approach fails, it may still be possible to recover the signal by using higher order derivatives of the sampled phase. More generally, procedures for the reconstruction of wrapped bandwidth-limited signals have been discussed in the signal processing literature, including the use of higher order derivatives [40].

Next, to evaluate the suitability of the data for coherent averaging calculations, we compare the full time-dependent phase to the sampled phase. As shown in Eq. (2), the phase of one RF comb line can be obtained from the heterodyne measurement setup using the data from one of the two cw lasers. We extract this phase and calculate the corresponding phase noise power spectral density (PN-PSD). The resulting PN-PSD is shown in Figure 6(a). The integrated RMS phase noise is also shown as a function of the lower limit of integration. The integrated noise starts to increase noticeably for lower integration limits below 10 kHz, reaching 1 rad at slightly below 1 kHz. In the context of the linewidth calculations of [41], we calculate that the frequency noise PSD crosses the so-called $\beta$ separation line at approximately 1.1 kHz, meaning that noise contributions above this frequency do not contribute to the linewidth of the RF comb lines. As in Figure 4, the increased noise at lower frequencies below 1 kHz is attributed to mechanical resonances. The individual beat notes between the cw lasers and the comb lines can also provide an upper bound on the width of the optical lines. Using the full duration of the measurement, we obtain an upper bound of about 2 MHz linewidth.

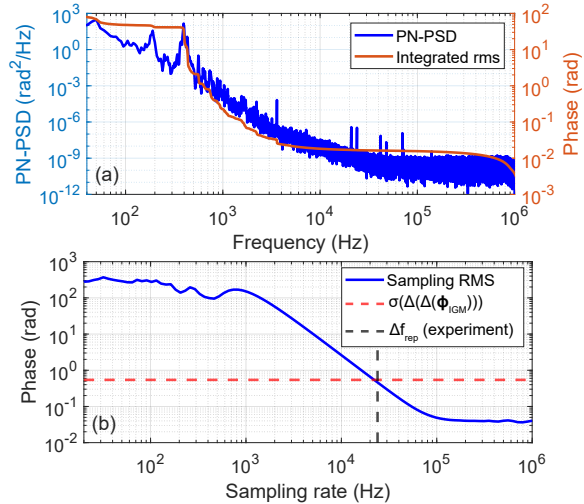

Fig. 6. Noise properties of one of the dual-comb lines in the RF domain obtained from heterodyne measurements between the combs and one of the cw lasers from Figure 1(c). Note, the influence of phase fluctuations of the cw laser is cancelled in this measurement. (a) One-sided phase noise power spectral density (PN-PSD) of the dual-comb line, and corresponding integrated phase. (b) RMS fluctuations of the interferogram phase as predicted by the PN-PSD (blue curve) and directly calculated (horizontal dashed line).

While useful to understand which spectral bands contribute to the phase noise, Figure 6(a) does not directly show us whether the sampled phase can be successfully unwrapped. To help evaluate the feasibility of this unwrapping task, consider the RMS fluctuations on samples of the IGM phase. In general, the RMS fluctuations on finite differences between samples of a time-dependent phase can be determined by a weighted integral of the phase noise power spectral density [42]. Consider a second-order finite difference of samples of a continuous variable $g(t)$, with constant sampling interval $\tau$. We write this as $\Delta_\tau^{(2)}(g)[n] \equiv g((n-1)\tau) + g((n+1)\tau) - 2g(n\tau)$, where $n$ is the array index. The RMS of this second-order phase-sample difference is equal to

$$\text{RMS}\left[\Delta_\tau^{(2)}(g)\right] = \sqrt{\int_0^\infty S_g(f)\left(2\cos(2\pi f\tau) - 2\right)^2 df}. \tag{8}$$

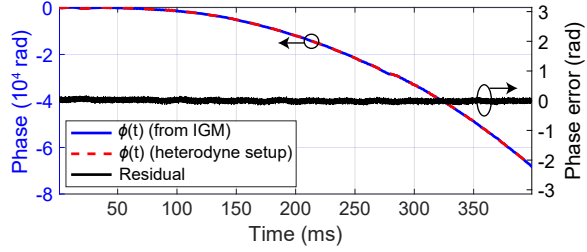

Fig. 7. Comparison between RF comb line phase obtained by cw heterodyne measurements to that obtained by interpolation of the sampled IGM phase.

When the RMS is too large (e.g. once it is $\gg 1$ rad) then it is more difficult to determine if an additional $2\pi$ phase slip has occurred between samples.

Using $S_\phi(f)$ from the beat note measurements (the PN-PSD of one RF comb line), in Figure 6(b) we plot $\text{RMS}[\Delta_\tau^{(2)}(\phi)]$ as a function of the sampling rate $1/\tau$. The horizontal dashed line indicates the RMS of the directly calculated second derivative of the IGM phase array $\phi_{\text{IGM}}$ from Figure 5, while the vertical dashed line indicates the actual repetition rate difference of the laser during the measurement. Since all three lines meet at the same point, the calculations are consistent. This implies that, at the sampling rate implicitly used in the measurement ($\tau = 1/\Delta f_{\text{rep}}$) the predicted RMS of $\Delta_\tau^{(2)}(\phi)$ from the PN-PSD (obtained by the cw-laser heterodyne data) agrees with the directly calculated RMS (obtained from the sampled and unwrapped IGM phase). Given these results, we see that the phase unwrapping algorithm described above would start to break down for repetition rate differences below 10 kHz, since the RMS would exceed $\pi$. Therefore, it is important to be able to access high repetition rate differences without aliasing. This condition is more straightforward to achieve with gigahertz lasers.

It should be noted that despite the high repetition rate difference of > 20 kHz, the main noise terms contributing to the fluctuations of the sampled phase are at frequencies below 1 kHz. The influence of these low-frequency fluctuations could be significantly mitigated by optimized optomechanical packaging of the laser. Moreover, it is possible to adapt the unwrapping algorithm discussed above to use a second-order phase derivative, which would reduce the influence of fluctuations below the sampling rate.

### 4.3. Suitability for coherent averaging

The calculation in Eq. (7) uses the full time-dependent phase of one RF comb line. Therefore, in order to implement a coherent averaging procedure using only the IGMs and no supplementary measurements, the values in the unwrapped phase array $\phi_{\text{IGM}}$ need to be interpolated between the IGM peaks (i.e. between the sampling points). In this subsection we implement this interpolation and compare the result to the phase directly determined by the heterodyne cw-laser measurements.

To obtain an IGM-based prediction of the phase of an RF comb line $\phi_{BN}(t)$, we first interpolate the IGM phase array $\phi_{\text{IGM}}$ on the time grid of the heterodyne beatnote measurements using a cubic spline method. This time grid is the one obtained via downsampling as discussed in section 3.1. After interpolation, we add a term corresponding to an integer number of RF comb lines multiplied by an integral of the repetition rate difference:

$$\phi_{BN}^{\text{predicted}}(t) = \phi_{\text{IGM}}^{\text{interp}}(t) + 2\pi n \int_{t_{\text{pk},0}}^{t} \Delta f_{\text{rep}}(t')dt'. \tag{9}$$

In this calculation, $\Delta f_{\text{rep}}$ is inferred from the IGM temporal centers $t_{\text{pk},j}$ and then interpolated on the beatnote time grid. To determine the integer $n$, consider the phase difference between (i) the heterodyne beatnote phase sampled at the temporal centers of the IGMs, and (ii) the directly calculated IGM phase:

$$\delta\phi_k = \phi_n(t_{\text{pk},k}) - \boldsymbol{\phi}_{\text{IGM}}[k]. \tag{10}$$

Comparing Eqs. (5) and (6), the $\Delta f_{\text{CEO}}$ and $f_{\text{rep}}$ terms are removed by this calculation, leaving only an integer number of cycles. Hence, in analogy to Eq. (4), the appropriate integer $n$ is given by $n = (\delta\phi_{i+k} - \delta\phi_i)/(2\pi k)$, which we insert into Eq. (9). The calculated and predicted phases are illustrated in Figure 7. Note, since the signals have a frequency of order 250 MHz, an overall linear part has been removed from both curves so that the relevant trends are visible on the graph. The black line shows the difference between the predicted and directly calculated phase, showing excellent agreement over the entire trace. This result validates that, using only the IGM data alone, we can predict the absolute phase of the RF comb lines to a precision of $\ll 1$ rad over arbitrary timescales and with no gaps.

## 5. Dual-comb spectroscopy

The laser is suitable for a wide range of experiments including pump-probe measurements, ranging, imaging, as well as linear and nonlinear spectroscopy techniques. For most linear absorption spectroscopy applications, an additional wavelength conversion step would be performed in order to access wavelength regions with richer spectroscopic signatures. However, such efforts would go beyond the scope of this paper. Therefore, here we perform a proof of principle dual-comb spectroscopy (DCS) measurement on an acetylene ($C_2H_2$) cell in order to validate the suitability of the source for spectroscopy measurements in general. We first discuss the coherent averaging procedure and then present the spectroscopy measurements.

### 5.1. Coherent averaging procedure

When dealing with real combs subject to timing and phase noise on the IGMs, Eq. (5) can be generalized to an integral over time. Application of Eq. (7) then yields a quantity whose IGMs always have the same phase at the peak, but the arrival times can still change due to fluctuations in $\Delta f_{\text{rep}}$. Such changes correspond to a time-domain stretching/compression of the electronic waveform and can be removed by adaptive scaling. This scaling amounts to interpolation of $U_n(t)$ onto a fixed optical delay grid, where the mapping from electronic times to optical delays is given by the (time-varying) ratio $\Delta f_{\text{rep}}/f_{\text{rep}}$. Thus, coherent averaging can be performed as follows:

1. Define an optical delay grid whose length is exactly $1/f_{\text{rep}}$
2. Determine $\phi_n(t)$ and $\Delta f_{\text{rep}}(t)$ (via the routines discussed in Section 4.2)
3. On each IGM period, extract a portion of the data slightly exceeding one period and apply Eq. (7), using a fixed value of index $n$ corresponding to a comb line near the center of the RF spectrum.
4. Interpolate $U_n(t)$ onto the fixed optical delay grid using the current value of $\Delta f_{\text{rep}}(t)$.
5. Average this waveform over all periods in the data.

The resulting averaged quantity corresponds to a Fourier series whose Fourier components are the RF comb lines. The true (and time-varying) frequencies of these RF lines can be reconstituted by Eq. (9) if needed, provided that the fluctuations were well resolved by the phase samples $\phi_{IGM}$. Note, in order for the underlying spectroscopic signal to remain unchanged, drifts in the optical comb lines of the beam sent through the sample should also be small compared to the width of the spectroscopic features under investigation within the averaging time.

It should be noted that our approach has several similarities to [36], since we use the same core approach of a time-domain analysis of the IGM phase and timing jitter to perform computational correction/averaging. Here, that approach is generalized to (i) provide a framework to evaluate the comb line resolvability based on a weighted integral of the phase noise power spectral density and/or IGM phase unwrapping; (ii) the modified algorithm described above, which begins with an exact retrieval and subsequent removal of the complete RF comb line phase before interpolation onto an optical delay grid; and (iii) experimental validation of the RF comb line phase retrieval by comparison to the heterodyne beat note measurement. The latter provides a stringent test for the comb line resolvability of a dual comb source (whether a pair of lasers or a single cavity solution), while (i) should be helpful as a basis for comparing different dual comb systems.

### 5.2. Spectroscopy measurements

Figure 8(a) shows our spectroscopy measurement where we installed an acetylene cell in the path of one of the two combs, then combined them on an uncoated YAG window before fiber coupling and a fiber-coupled photodiode. The acetylene cell had a nominal length of 20 cm and pressure of 987 mbar and was at room temperature. The total power reaching the photodiode (Thorlabs DET01CFC) was 0.9 mW (sum of both combs). The saved oscilloscope trace covers a time window of 0.8 s, corresponding to 16897 interferograms. Using the procedure of Section 5.1 we coherently average all of these interferograms to obtain the time-domain free-induction decay signal. The Fourier transform of this signal is the transmitted spectrum as a function of optical frequency shift. The only additional step needed is a shift in the frequency axis. The appropriate frequency shift can be determined by a fit to the reference data from HITRAN [43]. In Figure 8(b) we convert the measured to absorbance and compare it to HITRAN (upper panel), showing good agreement. The differences between the two (see residuals on the lower panel) are within the noise floor of the measurement. The noise on these residuals increases towards lower wavelengths because this goes further onto the wings of the optical spectrum. These measurements validate that we can perform coherently averaged DCS with the laser.

## 6. Discussion and Conclusions

We have demonstrated a single-cavity dual-comb source providing high average power (up to 3 W), high repetition rate (slightly above 1 GHz), and short pulses (as low as 78 fs). The laser is modelocked using a SESAM and is reliably self-starting. We have demonstrated dual-comb modelocking through use of a Fresnel biprism in transmission for the first time. Operation in transmission allows for convenient tuning without the need to move any reflective optics, and enables a pair of closed optical cavity paths in the short Z-shaped cavity with only one curved mirror. The high laser repetition rate in turn enables a large repetition rate difference without aliasing.

In contrast to fiber lasers, solid state lasers enable higher repetition rates, higher power, lower noise from RIN and quantum noise contributions, and can achieve this without any amplification or pulse compression. One of the concerns with solid-state lasers is complexity, but our approach overcomes this by using a single cavity with a single pump diode and a free-running system. Our approach also has several compelling advantages compared to other high power single-cavity 1-µm lasers. Modelocked thin-disk lasers can offer much higher average power than bulk-crystal DPSSLs [44], and have been used for dual-comb generation [25, 45]. However, they are larger and more complicated, cannot readily access gigahertz repetition rates, and they are susceptible to significant timing jitter contributions from water cooling of the reflective thin-disk. Thus, the novelty of our approach is that it combines multi-watt average power with femtosecond pulses, high repetition rate, high coherence, and a simple compact laser cavity.

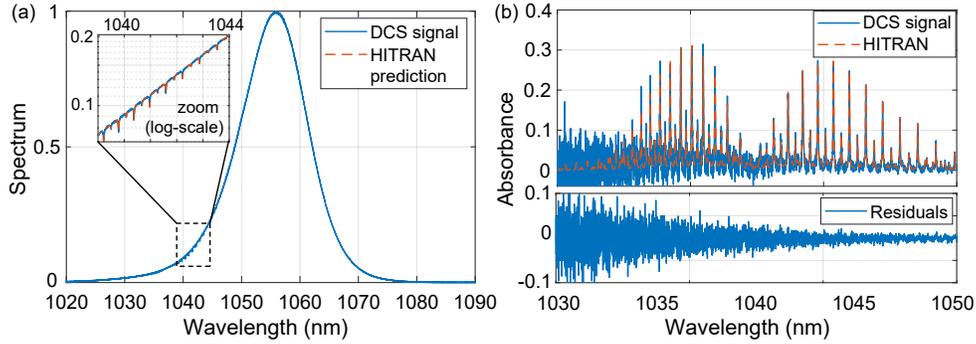

Fig. 8. Coherent dual-comb spectroscopy on acetylene using free-running gigahertz dual-comb laser. The repetition rate difference was 21.12 kHz and the measurement time was 0.8 s, corresponding to 16897 interferograms. All the interferogram periods are coherently averaged with the procedure discussed in Section 5.1. (a) Magnitude of the Fourier transform of the averaged interferograms; each point corresponds to the product between the electric field coefficient of a line of comb 1 and a line of comb 2. A frequency offset has been applied to line up the observed absorption features with the expected ones from HITRAN. Inset: zoom to a region containing absorption features. (b) Absorbance as a function of wavelength around the main absorption features. The power transmission is $\exp(-A)$ for absorbance $A$. The HITRAN curve assumes a cell length of 21 mm (which is within the physical tolerances of the component) together with absorption cross section data from HITRAN. The corresponding residuals are shown in the lower panel.

We characterized the laser's noise properties including timing and phase fluctuations. The pulse timing fluctuations of the two combs are highly correlated, which we validated by comparing conventional timing jitter measurements to the uncorrelated timing jitter obtained via heterodyne beat notes (> 10 dB noise suppression at high frequencies, and far more in the 100 to 1000 Hz range). We also looked in detail at resolving and tracking the fluctuations of the dual-comb lines in the radio frequency spectrum. This analysis yielded a key finding: by obtaining low laser noise and operating the system at a high repetition rate difference, the fluctuations of any RF comb line can be tracked simply by measuring the dual-comb interferograms. Consequently, coherent averaging of any dual-comb measurement using this free-running laser can be performed, without loss of information, over long timescales. High frequency noise contributions, which are typically caused by relative intensity noise [20, 21], have a negligible influence and do not compromise the coherence of the RF comb lines. Further, since the dominant noise contributions likely originate from mechanical resonances in combination with acoustic noise and vibrations, the noise could be suppressed even further by use of an integrated laser housing. Note, for cases where precise control of the optical frequencies themselves is needed, one could control fluctuations in $f_{\text{rep}}$ and $f_{\text{CEO}}$ of one of the combs. However, this is often unnecessary when comparing the measurements to known spectroscopic features (as we did here for acetylene), and foregoing such control simplifies the system significantly. In this paper we work with saved oscilloscope traces, but the spectroscopy algorithm can be adapted for real-time data processing by advanced data acquisition cards or a field-programmable gate array (FPGA).

The high repetition rate enables fast measurements while still offering relatively high (GHz)

resolution in the optical domain. This supports either high sensitivity (when averaging thousands of interferograms) or real-time DCS measurements (with limited or no averaging). The resolution is sufficient to measure a wide variety of gas species in ambient pressure conditions. As a proof of principle experiment, we measured transmission of an acetylene cell (987 mbar, 20 cm). The coherently-averaged measurement shows well resolved features in excellent agreement with reference data. Compatibility with coherently averaged DCS also implies the source is suitable for pump-probe applications as well, since these have more relaxed timing noise requirements.

While the 1-µm wavelength band is relevant for nonlinear spectroscopy [46, 47], it is of limited interest for linear gas spectroscopy since few targets absorb there. Similar to other 1-µm dual-comb results, we have used acetylene as a proof of principle experiment to illustrate the suitability of the source for spectroscopy measurements *in general*. For practical DCS applications one would first convert the wavelength to a band exhibiting strong absorption for the target of interest, for example in the mid-infrared. The substantial pulse energy of 3-nJ and short 80-fs pulses make the source very well-suited for driving various nonlinear processes directly from the oscillator output. These include coherent supercontinuum generation, mid-infrared difference frequency generation, and optical parametric oscillators (OPOs). Furthermore, the high power per comb line (up to 100 µW) together with the compact experimental setup makes the source promising for achieving high sensitivity in DCS. Combining the source with an efficient frequency conversion stage such as a single-cavity dual-comb OPO [48] thus represents a promising path towards shot-noise-limited multi-species spectroscopy in the mid-infrared. More generally, high power gigahertz single-cavity dual-comb lasers will be versatile tools for diverse applications including absorption spectroscopy from the terahertz to visible spectral range, pump-probe sampling, linear and nonlinear microscopy methods, and high-precision laser ranging.

**Funding.** Swiss National Science Foundation (40B2-0_180933, 40B1-0_203709); European Research Council (966718)

**Disclosures.** The authors declare no conflicts of interest.

**Data availability.** Data underlying the results presented in this paper are available in the ETH Zurich Research Collection, www.research-collection.ethz.ch.